\documentclass[10pt]{article}

\usepackage[utf8]{inputenc} 
\usepackage{textcomp}
\usepackage{graphicx}

\usepackage[affil-it]{authblk}
\usepackage[title,titletoc,toc]{appendix}
\usepackage{cite} 
\usepackage{url}  
\usepackage{ifthen}  

\usepackage{caption}
\usepackage{color}

\usepackage[cmex10]{amsmath}

\providecommand{\keywords}[1]{\textbf{\textit{Keywords---}} #1}

\usepackage[intoc]{nomencl}

\makenomenclature

\begin{document}

\title{Modelling Computational Resources for Next Generation Sequencing Bioinformatics Analysis of 16S rRNA Samples}

\author{Matthew~J.~Wade\thanks{Electronic address: \texttt{matthew.wade@ncl.ac.uk}; Corresponding author}}
\author{Thomas~P.~Curtis}
\author{Russell~J.~Davenport}
\affil{School of Civil Engineering and Geosciences, Newcastle University, Newcastle-upon-Tyne, NE1 7RU, UK}

\maketitle

\begin{abstract} 
In the rapidly evolving domain of next generation sequencing and bioinformatics analysis, data generation is one aspect that is increasing at a concomitant rate. The burden associated with processing large amounts of sequencing data has emphasised the need to allocate sufficient computing resources to complete analyses in the shortest possible time with manageable and predictable costs. A novel method for predicting time to completion for a popular bioinformatics software (QIIME), was developed using key variables characteristic of the input data assumed to impact processing time. Multiple Linear Regression models were developed to determine run time for two denoising algorithms and a general bioinformatics pipeline. The models were able to accurately predict clock time for denoising sequences from a naturally assembled community dataset, but not an artificial community. Speedup and efficiency tests for AmpliconNoise also highlighted that caution was needed when allocating resources for parallel processing of data. Accurate modelling of computational processing time using easily measurable predictors can assist NGS analysts in determining resource requirements for bioinformatics software and pipelines. Whilst demonstrated on a specific group of scripts, the methodology can be extended to encompass other packages running on multiple architectures, either in parallel or sequentially.
\end{abstract}
  \bigskip

\section{Introduction}
\subsection*{Next-Generation Sequencing Analysis}
The rapid increase in the utilisation of Next Generation Sequencing (NGS) technologies amongst disparate fields of research has resulted in a new set of challenges for scientific researchers. Whilst technology selection, sequencing costs and efficacy of sample preparation are still considerable concerns, pragmatic approaches in decision-making and implementation will result in satisfactory sequence generation. However, potentially vast quantity of data generated by sequencing efforts suggests that the true bottleneck is the computational analysis of the sequence data\cite{scholz12, lin07}. Indeed, without considered planning of the bioinformatics analysis, data processing time and costs can far exceed those of the actual sequencing itself \cite{angiuoli11}, diminishing the benefits attributed to high-throughput technologies.\par
Whilst the utilisation of computing resources and data analysis tools is generally accessible to the scientific and research communities, the ability to harness their full potential is often limited to specialists. Faced with the increasing ubiquity of bioinformatics tools for post-sequencing analysis and the realisation that decentralised research is becoming more prevalent, bioinformaticians are tasked with creating applications that are reliable, scalable, and user-friendly.\par
The prevalence of bioinformatics workflows and pipelines such as Galaxy\cite{goecks10}, QIIME core analysis\cite{caporaso10} and Taverna\cite{hull06}, and the increasing collaborative efforts through shared infrastructures\cite{ragan-kelley13} suggests greater uptake by non-specialists will be forthcoming. As NGS technology develops, the challenges faced by both bioinformaticians and users relate specifically to the competency of the software tools and the performance of the hardware to handle increasingly larger and more complex datasets. \par
The lack of appropriate hardware infrastructure is the greatest contributing factor to the bioinformatics bottleneck and the rise in virtual environments, parallelised code and super-computing facilities is testament to an understanding of the need for continual development and innovation in NGS data handling and management\cite{schatz10}. However, these structural and programmatic facilitators are not without their drawbacks. For example, cloud computing facilities such as Amazon Web Services' Elastic Compute Cloud \cite{aws13} offer flexible and scalable environments for performing a wide range of bioinformatics, but issues around data security and file transfer rates coupled with per hour usage costs make resource planning an integral requirement for any data analysis project.

\subsection{Understanding capacity and performance}
The capacity for processing and analysing NGS data accurately is dependent on identifying the most suitable software and hardware for the task. Incorrect selection or mismatch between data requirements and architecture will inevitably lead to suboptimal performance and, potentially, poor or erroneous results. With a wide range of bioinformatics tools being utilised by both specialists and non-specialists, there is a need for greater transparency in their deployment to facilitate effective and efficient analysis. As time and cost are often constraining factors in research and corporate environments, the ability to assign resources with \textit{a priori} knowledge of the performance and run time is of great benefit. For example, \textit{Cunningham} was developed to provide accurate runtime estimates for BLAST analysis of large shotgun sequence datasets \cite{white11} .\par
A recent study using the 16S gene for estimating bacterial diversity has shown the quantity and size of sequence clusters affects accuracy in non-parametric diversity calculations, whilst also determining which methods to employ\cite{barriuso11}.\par
Parallelisation of bioinformatics algorithms aimed at dramatically decreasing their processing time by exploiting multiple core processors\cite{galvez10} or GPU capabilities\cite{liu12} has alleviated some of the analysis bottleneck. Code optimisation can also make significant performance gains in highly parallel applications\cite{tan06}, but often requires expertise in coding that is not always accessible or practical for the end-user.\par
Parallel speedup and efficiency are key performance metrics that can be used to assess the most effective use of multiple CPU cores or nodes in a Cluster, Grid or Cloud environment. Whilst it may be intuitive that splitting large computational jobs amongst a greater number of processors should lead to increasing reductions in processing time, the presence of code that must be run serially (\textit{ser}) in most algorithms means that only a fraction of the work benefits from parallel (\textit{par}) speedup. Amdahl's Law\cite{amdahl67} describes the speedup of a process across multiple cores ($P$) given an amount of work ($N$) as:

\begin{equation} \label{eq:speed}
S(N,P) = \frac{t(N, P=1)}{t(N,P)}
\end{equation}\par

In ideal parallel processes, speedup is therefore equal to $1/P$, but with a fraction of code being serial, this equation becomes:

\begin{equation} \label{eq:speed_ser}
S(N,P) = \frac{t(N)_{ser} + t(N)_{par}}{t(N)_{ser} + \frac{t(N)_{par}}{P}}
\end{equation}\par

Knowing the speedup of parallelisation, then the efficiency may also be calculated simply by:

\begin{equation} \label{eq:eff}
E = \frac{S(N,P)}{P}
\end{equation}\par

For algorithms that are parallelisable, it is useful to perform these calculations to get an understanding of the scalability of the processes on any given architecture. This will help users to more appropriately assign resources and avoid problems with latency or parallel overheads.

\subsection{Computational transparency for targeted analysis}
Recent advances in sequencing technology have brought about unprecedented resolution in identification and classification of bacterial species. Sequencing of the highly conserved 16S rRNA gene is popular as it allows for comparative studies of microbial communities, their diversity and structure. The ubiquity of this approach in amplicon-based metagenomics coupled with the dwindling cost of high-throughput sequencing has put emphasis on development of tools and hardware infrastructure that can handle increasingly data rich sequence analysis.\par
A discretised pipeline was developed to model the relationship between sequence data size and complexity, and computational resource. An overview of the pipeline components is shown in Fig.~\ref{fig:PP} (Appendix~\ref{suppmat}), comprising several typical processing and analysis protocols available in the QIIME software. The performance was measured using the time taken to complete each process step in real and CPU metrics. The clock or \textit{real} time is necessary to determine actual resource cost, but can be skewed on systems where other extraneous processes are running, adding load to the shared resource. CPU time, characterised as the sum of \textit{user} and \textit{system} time, is reflective of the actual work done by the process being monitored. 

\section{Methods}
\subsection{System architecture}
The system used for evaluation was a 64-bit 2 x 6-Core (Intel Xeon 2.66 GHz CPU) Apple MacPro with 32GB RAM running OS X 10.8.2.\par
For the performance testing 16 logical cores (8 physical cores with hyper-threading) were used for the parallelisable components of the pipeline (e.g. Denoising) and single CPU otherwise. All analysis results are specific to this architecture and configuration. All cited instances of QIIME relate to its OSX compilation, MacQIIME Version 1.6.0.
 
\subsection{Training and validation datasets}
\subsubsection{Training data}
Three datasets containing 16S microbial rRNA gene fragments were used to develop the performance models. The first training dataset (\textit{MFC}) was taken from an acetate fed microbial fuel cell reactor inoculated with arctic soil sourced from Arctic soil (Ny-\r{A}lesund, Spitsbergen, Svalbard), and operated at 26.5\textdegree C \cite{heidrich12}. The second dataset was generated from a sample taken from a small eutrophic lake in the English Lake District (\textit{Priest Pot}) in 2008 \cite{quince09}. The third dataset (\textit{Arctic}) were sequences derived from DNA extracted from Arctic mineral soil samples collected from the Svalbard region [Unpublished].\par
The Priest Pot sequences were generated using standard 454 GS-FLX chemistry and targeted the V5 hyper-variable region, whilst the MFC and Arctic samples were sequenced with the more recent GS-FLX Titanium chemistry, which gives longer read lengths and targeted the V4-V5 regions, as at the time of sequencing (2011), they provided the highest classification accuracy with lowest amplification bias.\par
In the case of the MFC and Arctic data, samples were originally pooled using barcodes to provide a multiplexed dataset. However, only one sample was selected and processed through the pipeline to avoid the effects of redundancy when processing multiple samples as a single batch. 
\subsubsection{Validation data}
The first validation dataset (\textit{Mix}) were sequences taken from a laboratory scale batch reactor sample, which had been used to study anaerobic digestion of domestic wastewater at 15\textdegree C. The sample was sequenced in January 2013 using the same method employed for the {\textit{Arctic}} data. The second validation dataset (\textit{Artificial}) consisted of an artificial community from 90 clones that was pyrosequenced over the V5 region of the 16S rRNA gene with a 454 GS-FLX sequencer\cite{quince09}. 

Table~\ref{tab:datasets} summarises each of the datasets used in this analysis and includes the number of reads, average read length, number of putative OTUs and $\alpha$ diversity (equitability) of the raw, unfiltered sequences. Because of inherent sequencing errors, the OTU and equitability values are likely to be over- and underestimated, respectively. As can be seen, the samples sequenced with the older 454 GS-FLX chemistry are much shorter than those sequenced with the Titanium chemistry. After trimming to remove primer and barcode, the mean read lengths are approximately 200 bp for GS-FLX and 400 bp for Titanium. 

\noindent
\begin{center}
\begin{table*}[ht]
\caption{Test and Validation Datasets: Number of Sequences (Seqs); Average Read Length (bp); Number of OTUs at 97\% Cutoff (OTU); Equitability Estimate \cite{krebs89} of Total Samples ($\alpha$)}
\label{tab:datasets}
\centering
      \begin{tabular}{|c|c|c|c|c|c|}
       \hline 
        Source & Dataset & Seqs  & bp & OTU & $\alpha$ \\ \hline\hline
        MFC & Test & 72003 & 411 $\pm$ 45 & 828 & 0.481\\ \hline
        Priest Pot & Test & 28361 & 244 $\pm$ 52 & 1146 & 0.613\\ \hline
        Arctic & Test & 21576  & 426 $\pm$ 53 & 2267 & 0.807\\ \hline
        Mixed sediment & Validation & 19718 & 423 $\pm$ 49 & 2390 & 0.801\\ \hline
	Artificial & Validation & 46341 & 260 $\pm$ 38 & 177 & 0.536\\ \hline
      \end{tabular}    
      \end{table*}
      \end{center}

\subsection{Analysis steps}

An initial subsampling of the test datasets was performed using the QIIME \textit{subsample\_fasta.py} script to randomly split the raw Fasta file into subsamples of 5\% to 95\% fractions at intervals of 5\%. Six repeats were generated at each interval for the GS-FLX sequenced Artificial dataset and two repeats for the Titanium sequenced Arctic dataset (due to the greatly increased computational time required to denoise these sequences). Thus, for the denoising steps, 76 samples were used for training and 133 samples for validation.  \par
A simple bash script was written to simplify and automate the performance testing by looping through the subsamples, processing according to each step in the performance pipeline (Fig.~\ref{fig:PP}) and passing the Real and CPU time for execution to a separate output file.\par
The subsampled Fasta files are pre-processed using the QIIME \textit{split\_libraries.py} script, which applies some basic quality filtering to the sequences for read length trimming, ambiguous base checking and primer and barcode removal.\par
The subsamples are then ready for denoising to correct for errors generated in the PCR and sequencing steps. QIIME Denoiser \cite{reeder10}  is a heuristic algorithm that uses a greedy alignment scheme before clustering flowgrams in descending order of abundance. Erroneous reads are filtered from the cluster to produce the final denoised sequences. Denoiser can use pre-filtered Fasta files and matches the IDs of the remaining reads with those present in the text translation of the raw SFF file, to avoid denoising of poor quality sequences. Chimera checking is an optional but often important step that is performed independently from denoising using the ChimeraSlayer\cite{haas11} tool via QIIME's \textit{identify\_chimeric\_seqs.py} script.  \par
For denoising using AmpliconNoise, the Standard Flowgram File (SFF) associated with the data was split into 19 subsamples with sizes corresponding to the set used for QIIME denoising using SFF Workbench\cite{heracle12} via a Wine\cite{wine13} translation of the Windows API. The individual SFF files were then converted to text translations using the QIIME \textit{process\_sff.py} script. There is no need for demultiplexing of the data as only a single sample is used in the pipeline test. The AmpliconNoise software\cite{quince09} uses Bayesian theory to generate an approximate likelihood from empirical error distribution data to infer true read identity given sequencing error (PyroNoise) and PCR error (SeqNoise). An additional chimera checking step using Perseus\cite{quince11} is performed after error removal. The software was run using the QIIME wrapper script \textit{ampliconnoise.py} rather than via the stand-alone package to maintain a consolidated workflow.\par
The denoised reads from QIIME denoiser were used for downstream analysis using the following steps:
\begin{itemize}
\item \textbf{De Novo OTU picking:} Clustering of sequences with 97\% similarity threshold using the \textit{uclust} method\cite{edgar10}, before picking representative Operational Taxonomic Units (OTUs) from each cluster by the sequence first assigned to a cluster.
\item \textbf{Assign Taxonomy:} Assignment of taxonomic identities to the OTUs using the curated Greengenes 16S rRNA gene database\cite{mcdonald12} with a na\"{i}ve Bayesian classifier, \textit{RDP}\cite{wang07}.
\item \textbf{Alignment:} Alignment of the sequences is necessary for comparative analysis, such as $\beta$ diversity, in which phylogenetic distances are used to understand differences in community composition from distinct samples. Alignment was performed using the \textit{PyNAST} method and pairwise clustering with \textit{uclust}\cite{edgar10} against a Greengenes database template. Gap only columns and highly variable regions within the alignment files are removed using a filtering step that makes use of Greengenes compatible Lanemask file for excluding these positions. Phylogenic relatedness of organisms within a sample may also be of interest and the creation of a phylogenetic tree using the \textit{FastTree 2.1.3}\cite{price10} method is performed after alignment. 
\item \textbf{Diversity analysis:} Diversity metrics are key outputs from the QIIME pipeline and are used to gain an quantitative understanding of the distribution and relatedness of organisms within a sample ($\alpha$ diversity) or between different samples ($\beta$ diversity). $\alpha$ diversity uses the abundance data stored in the OTU table generated from the OTU picking and taxonomy assignment steps to calculate a range of metrics provided by the user, such as Chao1, Shannon and Phylogenetic Distance. Rarefaction plots are generated for each metric based on random subsampling (using a pseudo random number generator) of the OTU table between a given range of sequences per sample and at a given step size. $\beta$ diversity uses both the information stored in the OTU table and a phylogenetic tree, if the phylogenetic metrics are calculated using Unifrac. In this analysis, both quantitative (weighted) and qualitative (unweighted) Unifrac metrics\cite{lozupone05} were calculated and Principal Coordinate Analysis plots generated to display the results.
\end{itemize}\par
Details of the parameters used for each analysis step are provided in Table~\ref{tab:pipeparam} (Appendix~\ref{suppmat}).

\subsection{Performance measures}
Both the wall clock and CPU (Usr + Sys) were recorded using the GNU time command, which allows for formatting of the output in Mac OSX, and stored as text files. Whilst the wall clock time indicates the amount of real time between execution and completion of a process, CPU time is more indicative of the computing effort required to run the process. However, when considering running large datasets through a bioinformatics pipeline, time to completion is the measure by which costs can be assessed.\par
File input size was determined based on the process step being analysed as shown in Table~\ref{tab:seqtype} (Appendix~\ref{suppmat}). The number of reads and mean read length per Fasta file were determined using the QIIME \textit{count\_seqs.py} script. Equitability was calculated using the QIIME $\alpha$ diversity metric script on the raw input fasta files supplied to the denoising algorithms. \par
The time and predictor (diversity, number of reads and read length) data was imported into Matlab\cite{matlab13} and a stepwise Multiple Linear Regression (MLR) was applied to fit a model between the explanatory variables and clock time. The MLR model takes the following generalised form: 
\begin{equation} \label{eq:mlr}
y_{i} = \beta_{0} + \beta_{1}x_{i,1} + \beta_{2}x_{i,2} + ... + \beta_{p}x_{i,p} + \epsilon_{i}
\end{equation}

\noindent where $y_{i}$ are the predictands, $x_{i,p}$ the predictors, $\beta_{p}$ the regression coefficients, and $\epsilon_{i}$ is the error term. The regression coefficients are the solution to the least squares estimation:

\begin{equation}  \label{eq:mlrcoeff}
\beta = [X^{T}X]^{-1}X^{T}y_{i}
\end{equation}

\noindent where $X$ is the matrix of regressor variables. The Matlab function \textit{LinearModel}, from the Statistics Toolbox (v8.2), was used to perform the stepwise regression using the polynomial form described in eq.~\ref{eq:mlr}. The starting model included the intercept, linear terms, interactions and power terms with interactions up to a factor of 4 for each explanatory variable. The algorithm uses forward and backward regression with the Sum of Squared Error (SSE) to add and remove terms from the model based on the p-values of the F-statistic with and without a potential term. The thresholds for adding and removing terms were p $<$ 0.05 and p $>$ 0.1, respectively. \par
A conventional method for assessing goodness-of-fit for linear regression models is to calculate the $R^{2}$ value or coefficient of determination. As the complexity of the polynomial increases (by adding more variables), the $R^{2}$ value will increase, which may result in a skewed confidence in the reliability of the model. The adjusted-$R^{2}$ value is used to address this issue:
\begin{equation} \label{eq:rsq}
R^{2}_{adj} = 1 - \frac{(\frac{\sum{(\hat{y_{i}} - \bar{y})^2}}{\sum{(y_{i} - \bar{y})^2}})(n - 1)}{n - p -1}
\end{equation} 

\noindent where $y_{i}$, $\bar{y}$ and $\hat{y_{i}}$  are the observed data, mean of the observed data and modelled predictands, respectively, $n$ is the number of observations and $p$ the number of regressors. \par
Parallel speedup and efficiency were calculated according to Amdahl's Law for AmpliconNoise based denoising of the MFC sample, as this step is observed to be the most computationally intensive part of the pipeline and is compiled to run in parallel. Denoising was performed using 1, 2, 4, 8, 16, 32, 64 and 128 physical cores, with 48GB RAM per node (six cores per node) on the DIAG resource \cite{diag13} for two sub-samples of the total sequences (20000 and 40000 reads).

\section{Results and discussion}
\subsection{Model development}
Performance modelling was initiated on the two test datasets using the standard QIIME pipeline tools for processing and analysis of 16S rRNA sequencing data. The wall clock (Real) and CPU time were recorded for each pipeline step and a model fit between these variables and input read number was made. For the denoising algorithms (QIIME denoiser and AmpliconNoise) a Multiple Linear Regression (MLR) model was deemed necessary as visualisation of the read number versus clock time highlighted that more than one explanatory variable was influencing the denoising time. Community diversity within a sample, or more specifically taxonomic rank-abundance, is known to influence computational effort of denoising \cite{reeder10}. \par
The equitability, or evenness, defines the homogeneity of species within a community, with higher values of the index indicating a highly even or homogeneous distribution of species. This diversity metric was used as a second explanatory variable in the MLR model as a candidate for determining clustering speed during denoising. Despite diversity calculation being a component of the QIIME pipeline, the parameter values used for the model were generated by calculating equitability from the raw input fasta files, prior to quality checking and noise removal. Although the value is not a true measure of diversity due to the presence of errors, it is assumed the error profile across all samples is equivalent and this will not affect the model.

\subsection{Modelling sequence denoising strategies}
\subsubsection{Training}
Denoising of high-throughput sequencing data with the most commonly used algorithms (QIIME denoiser and AmpliconNoise) is clearly the major performance bottleneck in the analysis pipeline, but also one of the most critical in terms of determination of more accurate OTU numbers and subsequent classification.\par
A MLR model was developed with two explanatory variables (number of reads ($\lambda$) and sample equitability ($\alpha$)) as predictors and wall clock time ($y$) as the response variable.\par
By simply observing the relationship between the explanatory and response variables, it is evident that a non-linear implementation of the MLR model is necessary. The introduction of power terms in the model is intended to reflect the curvlinear nature of the underlying dependencies.\par
Stepwise MLR models were developed using the three training datasets for the QIIME denoiser (Eq.~\ref{eq:mlrqiime}) and AmpliconNoise algorithms (Eq.~\ref{eq:mlran}). The models take the form given by equation~\ref{eq:mlr} with non-linear power terms and cross-products between the two predictor variables.
\begin{multline}  \label{eq:mlrqiime}
y_{qd} = \beta_{1}\alpha + \beta_{2}\alpha\lambda + \beta_{3}\lambda^{2} + \beta_{4}\alpha^{2}\lambda + \beta_{5}\alpha\lambda^{2} + \beta_{6}\lambda^{3} + \beta_{7}\alpha^{2}\lambda^{2} + \beta_{8}\alpha\lambda^{3}
\end{multline}

\begin{figure}[t!]
\centering
\includegraphics[trim=0cm 0cm 0cm 0cm,clip=true,width=4.4in]{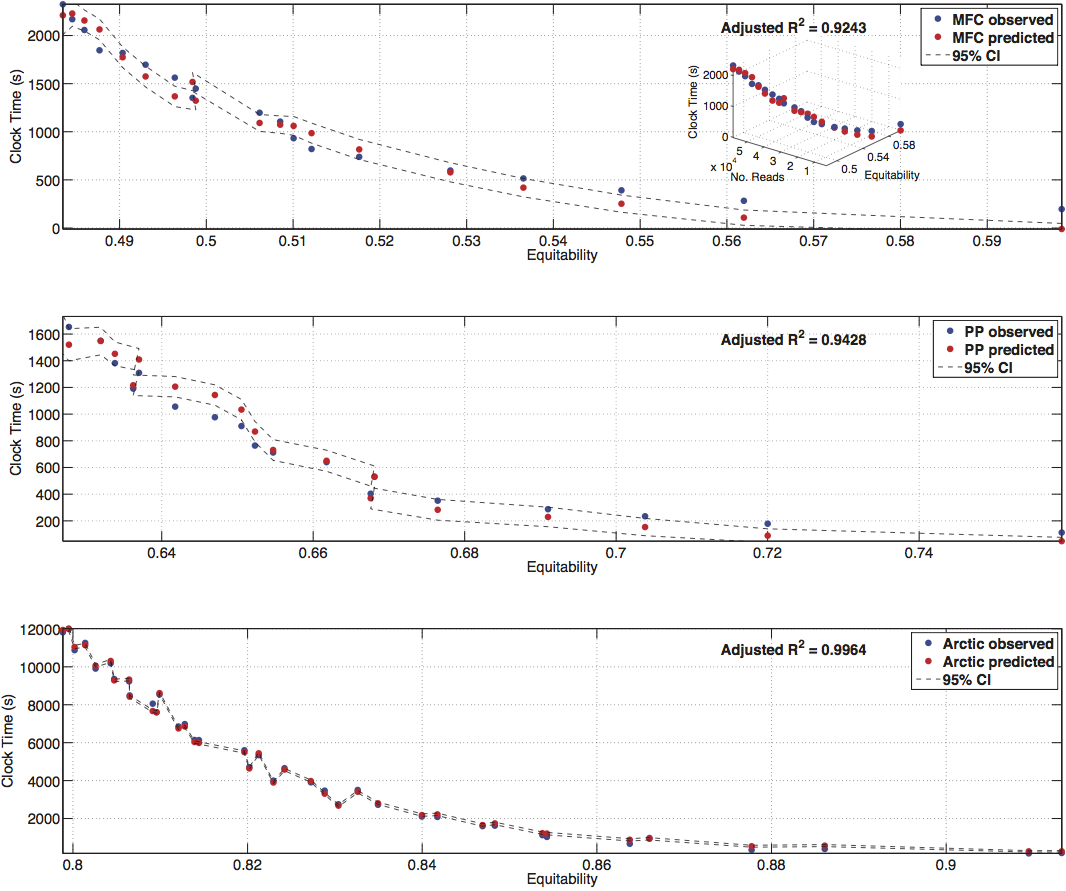}
\caption{QIIME denoiser model performance for the three training datasets; fit plotted along the equitability axis. Two parameter fit is shown in the 3D embedded figure}
\label{fig:calmodel_q}
\end{figure}

The regression coefficients ($\beta$) are shown in Table~\ref{tab:modelcoeff}. The results from the QIIME denoiser model suggest a conformity between the two explanatory variables selected and the resulting predictand. Fig.~\ref{fig:calmodel_q} shows excellent prediction (Adjusted $R^{2} >$ 0.9) for all training data, which is confirmed by performing an ANOVA on the full model (F-statistic = 7.38$\times 10^{3}$, p-value = 2.54$\times 10^{-95}$) indicating that the non-linear model is highly significant. All plots are shown in relation to equitability for ease of visualisation, however, an example 3D plot (embedded in the uppermost plot in Fig.~\ref{fig:calmodel_q}) for the MFC data shows the excellent fit against both explanatory variables.\par

\begin{figure}[t!]
\centering
\includegraphics[trim=0cm 0cm 0cm 0cm,clip=true,width=4.4in]{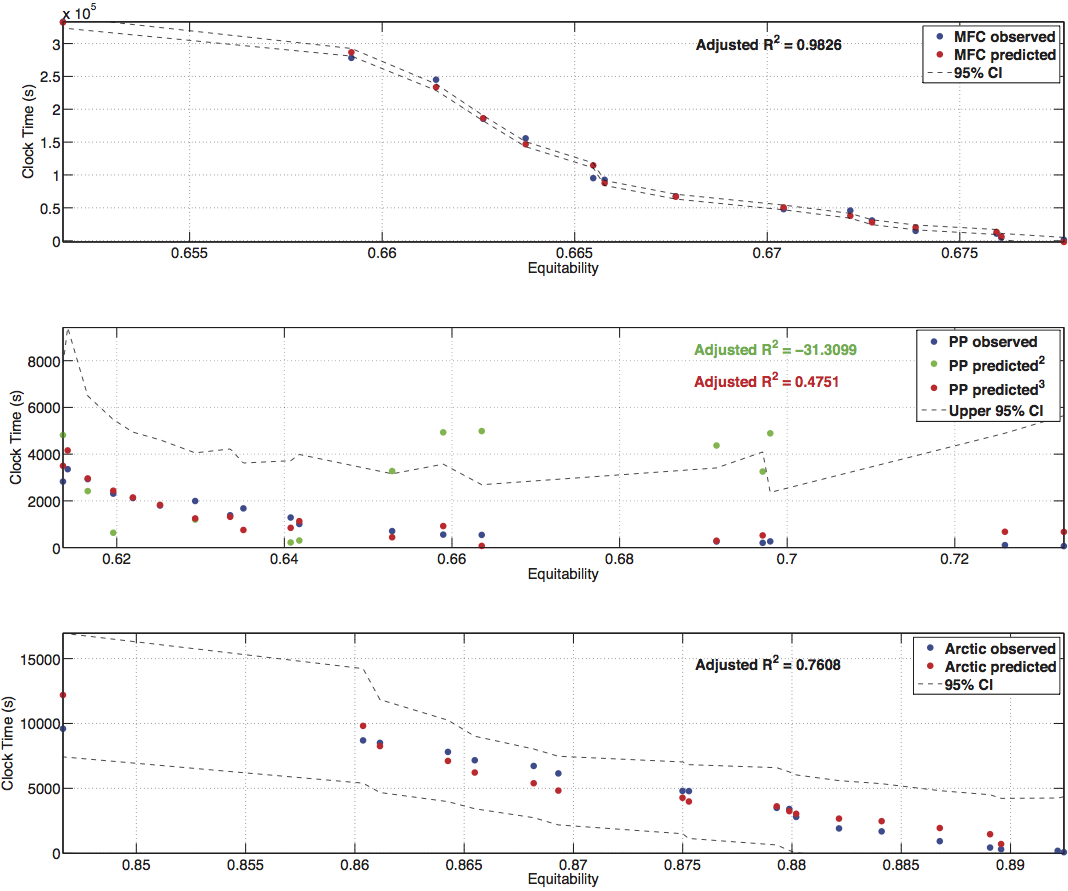}
\caption{AmpliconNoise model performance for the three training datasets; fit plotted along the equitability axis}
\label{fig:calmodel_a}
\end{figure}

For AmpliconNoise, an initial two parameter training model produced good fits for two datasets (Arctic and MFC), but could not fit the Priest Pot data. It was surmised that read length could be an additional factor in determining processing time during the sequencing error removal (\textit{Seqnoise}) step, given the importance of sequence size in influencing error rate distribution \cite{gilles11}. Including mean read length per sample ($\rho$) as a third parameter in the model decreased the Root Mean Square Error of Calibration (RMSEC) from 5750 to 129 and the improvement can also be seen in Fig.~\ref{fig:calmodel_a}. Although prediction is not as convincing as with the QIIME denoiser data, the model, shown in equation~\ref{eq:mlran}, is still highly significant (F-statistic = 1.60$\times 10^{3}$, p-value = 2.60$\times 10^{-51}$).
\begin{multline}  \label{eq:mlran}
y_{an} = \beta_{1}\rho + \beta_{2}\lambda + \beta_{3}\alpha\lambda + \beta_{4}\lambda^{2} + \beta_{5}\alpha\rho + \beta_{6}\rho\lambda + \beta_{7}\rho^{2}\\ + \beta_{8}\alpha\lambda^{2} + \beta_{9}\lambda^{3} + \beta_{10}\alpha\lambda^{3}
\end{multline}

\subsubsection{Validation}

\begin{figure}[!t]
\centering
\includegraphics[trim=0cm 0cm 0cm 0cm,clip=true,width=4.4in]{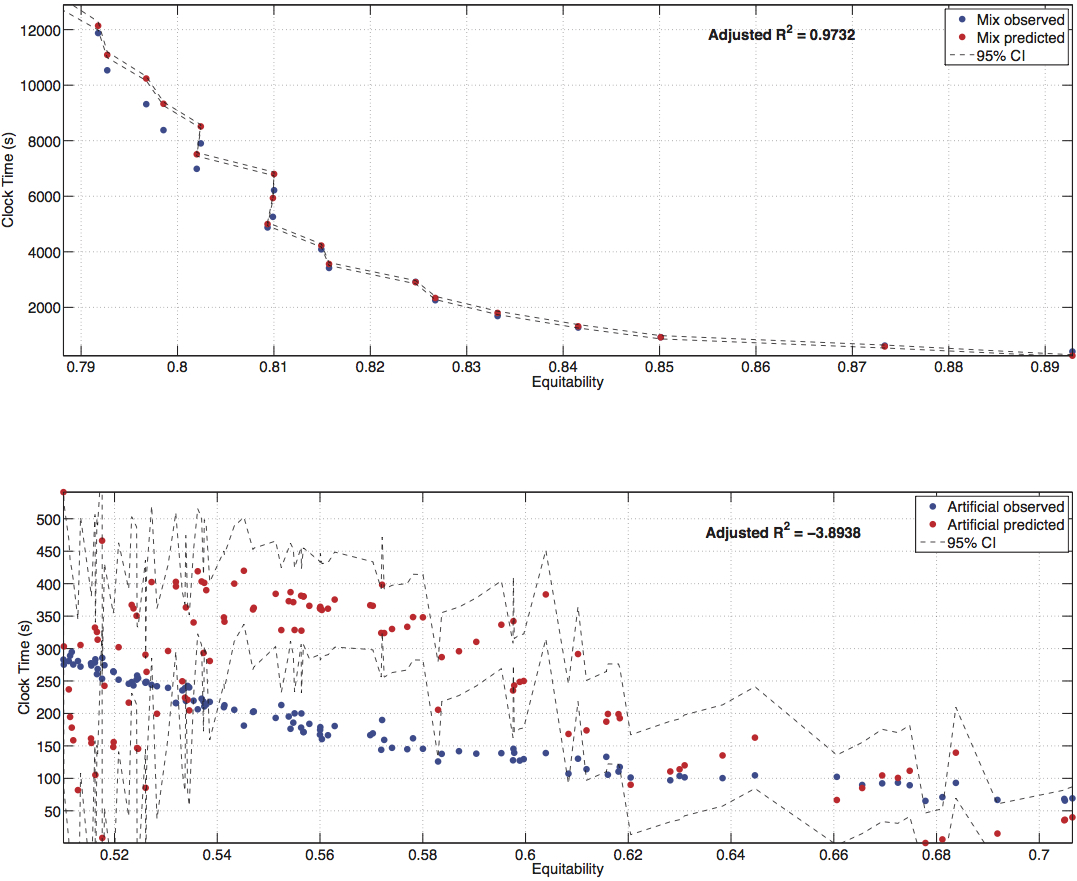}
\caption{QIIME denoiser model prediction for the two validation datasets; fit plotted along the equitability axis}
\label{fig:valmodel_q}
\end{figure}

Validation of regression based models is critical to ascertain their ability to be used with independent data. In complex, highly parameterised models, there is a risk that overfitting of the data may occur, in which the model tends to fit to the training data, but lacks the predictive capacity when fitting to validation or real-world datasets.
The models developed for both denoising algorithms were tested with the independent validation datasets to assess their suitability for  prediction of processing time. For both denoising algorithms the models fit the Mixed sediment dataset well, with adjusted $R^{2}$ values of 0.97 and 0.72 for QIIME denoiser and AmpliconNoise, respectively. However, the models do not predict the Artificial data (See Fig.~\ref{fig:valmodel_q}), which suggests that there is some underlying property of the artificially generated sequence communities that is not captured during training. The dataset was constructed to represent a community with log-normal distribution analogous to true community distributions found in the environment\cite{quince09}, and it is possible that this artificial construct has presented some feature in the data that has a significant impact on denoising performance. As discussed, the Priest Pot data was acquired from the same sequencing technology and the inclusion of the mean read length in the AmpliconNoise model had significant impact on training performance. \par

A main effects plot, shown in Fig.~\ref{fig:maineff} (Appendix~\ref{suppmat}), was generated to look at the contributions from each independent model variable on the clock time. For both the two and three variable models, the number of reads was the largest contributing factor to the output variability. It can also be seen that, although mean read length has a small impact on the AmpliconNoise model, justifying its inclusion, it is not the critical factor in explaining the poor performance with the Artificial dataset.

\subsection{General pipeline model}
The pipeline algorithms deployed sequentially without parallelism (i.e. on a single CPU core) generally contribute insignificant burden to the overall processing time compared to the denoising step. However, based on system memory availability and CPU processor speed, scripts related to sequence alignment, OTU picking, taxonomy assignment and diversity calculation, may become cumbersome, especially for large sequencing runs. QIIME includes several parallel commands to handle such conditions, but in this study the test environment and dataset sizes were such that single CPU processing was sufficient. A MLR model was developed using the total wall clock time measured for all analysis steps independent of denoising, as shown in Table~\ref{tab:seqtype}. Although non-continuous, as quality filtering occurs prior to denoising, whereas all other steps are downstream, the intention is to indicate a relationship between processing time and predictor variables that will aid resource allocation. Additionally, it should be noted that the predictor variables will undergo changes during the pipeline as reads are removed, trimmed and truncated, particularly at the upstream end of the pipeline. However, monitoring changes in read number and diversity is impractical, especially with automated pipelines and given the underlying aim of assessing resource requirements \textit{a priori}. It is assumed that the underlying correlation between predictors and response hold true for the agglomerated model as the relative changes in predictor variables are expected to be uniform across all training samples for any given pipeline. \par
\begin{figure}[!t]
\centering
\includegraphics[trim=0cm 0cm 0cm 0cm,clip=true,width=4.4in]{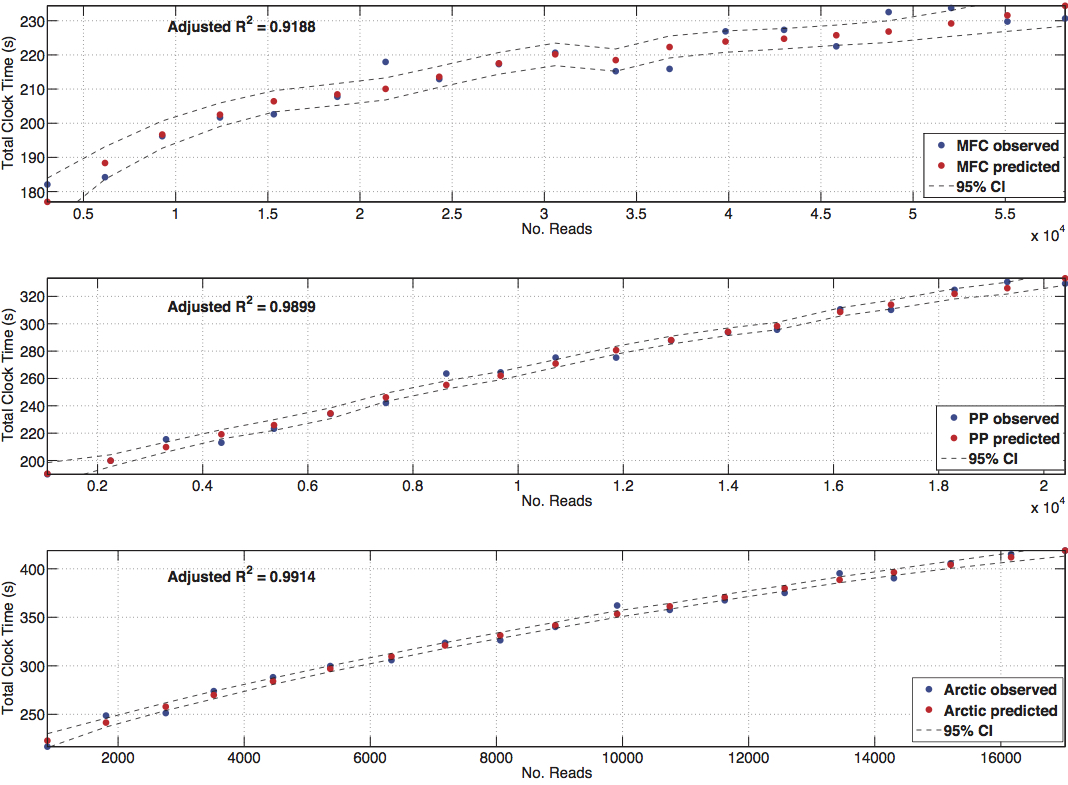}
\caption{General pipeline model performance for the three training datasets; fit plotted along the read number axis}
\label{fig:calmodel_p}
\end{figure}

\begin{figure}[!t]
\centering
\includegraphics[trim=0cm 0cm 0cm 0cm,clip=true,width=4.4in]{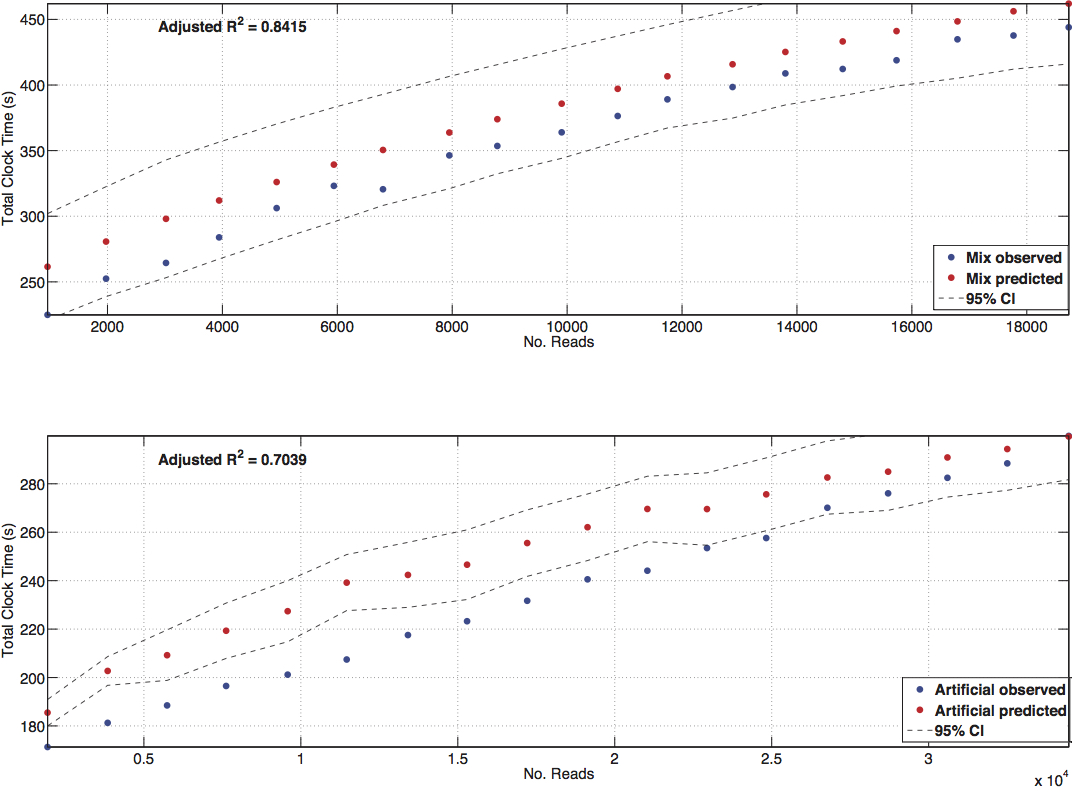}
\caption{General pipeline model prediction for the two validation datasets; fit plotted along the read number axis}
\label{fig:valmodel_p}
\end{figure}

 
Observation of individual pipeline step results indicated generally linear relationships between number of reads and clock time but, as with the denoising algorithms, additional confounding factors appeared to have a role in determining response across datasets. An initial model was developed using number of reads and equitability, which gave satisfactory training results but produced poor fitting for the Artificial validation data (Adjusted $R^{2}$ of 0.333). A three factor model that included mean read length was investigated and greatly improved prediction of the articitial validation dataset (Adjusted $R^{2}$ of 0.704), but with a slight decrease in the model fit for the Mixed dataset (Adjusted $R^{2}$ reduced from 0.914 to 0.841). The final pipeline MLR model (F-statistic = 1.27$\times 10^{3}$ , p-value = 5.43$\times 10^{-53}$) is shown in equation~\ref{eq:mlrpipe} and the regression coefficients in Table~\ref{tab:modelcoeff}, with the fitted training and prediction curves presented in Figs.~\ref{fig:calmodel_p} and~\ref{fig:valmodel_p}, respectively.
\begin{multline} \label{eq:mlrpipe}
y_{pipe} = \beta_{0} + \beta_{1}\lambda + \beta_{2}\alpha + \beta_{3}\rho + \beta_{4}\alpha\lambda + \beta_{5}\lambda^{2} + \beta_{6}\alpha\rho + \beta_{7}\lambda\rho + \beta_{8}\rho^{2}\\ + \beta_{9}\alpha\lambda^{2}
\end{multline}

\subsection{Speedup and Efficiency}
Due to the unfeasibly long processing time for AmpliconNoise and, to a lesser extent, QIIME denoiser, there may be a tendency to invest heavily in high-performance computing solutions to dramatically reduce the run time. Whilst there is some guidance on memory requirements for running the algorithms (1GB for FLX sequences with QIIME denoiser \cite{qiime13},  $>$ 8GB for large datasets when running AmpliconNoise \cite{anoise13}), there may be a tendency for employing as large a resource as economically and logistically feasible. Aside from the costs involved, acquisition of hardware for long-term use may result in redundancy unless demand is significant. Deployment on decentralised systems may also result in conflict if non-essential capacity is being utilised for the denoising task. \par
\begin{figure}[!h]
\centering
\includegraphics[trim=0cm 0cm 0cm 0cm,clip=true,width=4.4in]{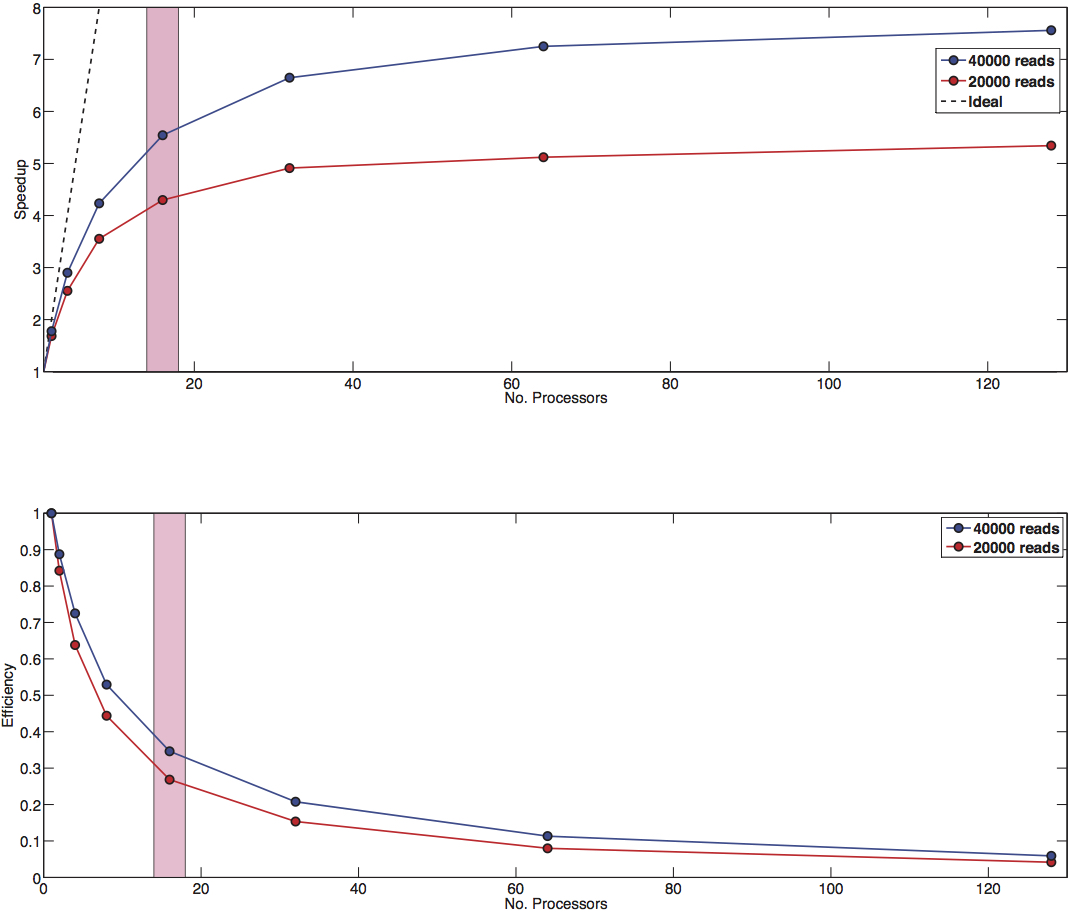}
\caption{Speedup and Efficiency metrics for subsampled MFC data (20000 and 40000 reads) across multiple cores on the DIAG resource using AmpliconNoise}
\label{fig:speedup}
\end{figure}

The results of the speedup and efficiency tests performed with AmpliconNoise on the decentralised DIAG \cite{diag13} resource are shown in Fig.~\ref{fig:speedup}. The speedup plot shows that actual performance improvement is far from ideal when utilising more processors, reaching a threshold of approximately 7.5 times speedup with 128 cores, which corresponds to 5\% efficiency shown in the second plot. AmpliconNoise is clearly not a massively parallel algorithm, with many serial components contributing to the dramatic reduction in efficiency with greater parallel resources.\par
Based on the analysis and considering the tradeoff between time to completion and resource utilisation/expenditure, between 8 and 16 (shown as a vertical bar in Fig.~\ref{fig:speedup}) cores appear reasonable for this architecture. Using more than 16 cores does not deliver significant increases in speedup, whilst efficiency drops below 30\%.

\begin{table}[!t] 
\caption{Regression Coefficients for the MLR Models of the QIIME Denoiser and AmpliconNoise Algorithms, and General QIIME Pipeline Steps}
\label{tab:modelcoeff}
\centering
      \begin{tabular}{|c|c|c|c|}
       \hline 
        Regress. coeff. & QIIME & Ampliconnoise & Pipeline\\ \hline\hline
        $\beta_{0}$ & 0 & 0 & 2820.200 \\ \hline
        $\beta_{1}$ & 1.718 & -47.687 & -0.012 \\ \hline
        $\beta_{2}$ & -5.360 & 1.709 & 556.420 \\ \hline
        $\beta_{3}$ & 1.1$\times 10^{-4}$ & -7.098 & -19.001 \\ \hline
        $\beta_{4}$ & 4.096 & 2.2$\times 10^{-4}$ & 0.045 \\ \hline
	$\beta_{5} $ & -4.6$\times 10^{-4}$ & 127.790 & 2.8$\times 10^{-7}$ \\ \hline
	$\beta_{6} $ & 6.2$\times 10^{-10}$ & 0.017 & -1.640 \\ \hline
	$\beta_{7} $ & 4.7$\times 10^{-4}$ & -0.184 & -2.1$\times 10^{-5}$ \\ \hline
	$\beta_{8} $ & -1.4$\times 10^{-9}$ & -5.2$\times 10^{-4}$ & 0.032 \\ \hline
	$\beta_{9} $ & ... & -1.6$\times 10^{-8}$ & -5.7$\times 10^{-7}$ \\ \hline
	$\beta_{10} $ & ... & 3.0$\times 10^{-8}$ & ... \\ \hline
      \end{tabular}    
      \end{table}
      
\section{Conclusion}
The QIIME software package is widely employed for microbial community analysis using data derived from a range of NGS technologies. The protocols defined by \cite{kuczynski11} offer a standard methodology for processing sequencing reads and transforming that data into interpretable information such a sample diversity and phylogenetic distances. The benefits of using QIIME for downstream analysis of NGS data is its ability to pipeline a range of bioinformatics steps in a consistent and reproducible manner, several parallelised scripts for data intensive processing and its portability across a range of high-throughput and high-performance environments such as Amazon Cloud, Virtual Box/CloVR and Grid services such as DIAG.\par
When considering what resources to utilise for post-sequencing analysis, it is important to have an understanding of computational requirements for the tools employed. This is vital when considering deploying algorithms on resources that provide a service at a cost related to time utilised. However, it may also be important for users requiring rapid turnaround from sequencing to information, those wishing to optimise a pipeline or invest in additional computational resources, and in cases where the resource is utilised by multiple analysts. \par
The work presented demonstrates that there exists a significant relationship between characteristics of the input sequencing data and the computational time required to process that data. Although the models developed are characteristic of the system that they were calibrated against, the MLR modelling technique can be applied to generate models for any architecture. \par
Two and three variable models developed for sequence denoising algorithms were successful in predicting time for completion, but suffered with the Artificial data. This was potentially due to the nature of the dataset construction, which was not captured in the model development, which consisted entirely of samples from naturally distributed communities. However, a three variable general pipeline model predicted total time for completion of the standard QIIME analyses with 6\% and 7\% error for the Mixed and Artificial datasets, respectively. \par
There is often a temptation when working with computationally intensive algorithms to allocate the maximum amount of resource to the problem without considering if this will be optimal. Parallel speedup and efficiency analysis for the AmpliconNoise algorithm revealed that denoising with increasing number of CPUs is far from ideal. The efficiency of the computation decreases rapidly beyond a single core processor, whilst speedup is not significantly increased beyond 32 cores and the analysis suggests the use of between 8 and 16 processors is sufficient under the test architecture. \par 
Accurate modelling of the relationship between input sequence data and time to completion has shown to be a viable method for supporting data analysts in decision-making related to resource allocation. Although restricted to a single architecture and set of tasks, the methodology can and should be applied in a broader context to understand if the approach is transferable across a range of applications (e.g. read mapping, genome assembly). A standardised model may be idealistic, but the modelling effort is minimal when contrasted with the expected data throughput that is beginning to emerge in sequencing today. As sequencing technology is evolving at a remarkable rate, it will be useful to look at assessing the methodology presented here against more diverse data sets, in terms of size and source environment. This should be coupled with attempts to identify how generic the models are across different system architectures and platforms. \par
It is suggested that by developing techniques and tools for modelling computational requirements for bioinformatics analysis, coupled with methods for estimating sequencing effort required to generate the necessary depth of information (see \cite{quince08}), then analysts can be armed with the capabilities to comprehensively plan their research efforts, assess resource requirements and funding needs. 

\nomenclature{MFC}{Microbial Fuel Cell}
\nomenclature{MLR}{Multiple Linear Regression}
\nomenclature{OTU}{Operational Taxonomic Unit}
\nomenclature{PCR}{Polymerase Chain Reaction}
\nomenclature{RMSEC}{Root Mean Square Error of Calibration}

\printnomenclature

\bibliography{Wade_IEEE_CB_B.bib}      

\begin{thebibliography}{10}

\bibitem{aws13}
{Amazon Web Services}.
\newblock Amazon elastic compute cloud (amazon ec2), March 2013.

\bibitem{amdahl67}
Gene Amdahl.
\newblock Validity of the single processor approach to achieving large-scale
  computing capabilities.
\newblock In {\em AFIPS Conference Proceedings}, volume~30, pages 483--485,
  1967.

\bibitem{angiuoli11}
Samuel~V Angiuoli, James~R White, Malcolm Matalka, Owen White, and W~Florian
  Fricke.
\newblock Resources and costs for microbial sequence analysis evaluated using
  virtual machines and cloud computing.
\newblock {\em PLoS One}, 6(10):e26624, 2011.

\bibitem{barriuso11}
Jorge Barriuso, Jose~R Valverde, and Rafael~P Mellado.
\newblock Estimation of bacterial diversity using next generation sequencing of
  16s rdna: a comparison of different workflows.
\newblock {\em BMC Bioinfomatics}, 12:473, 2011.

\bibitem{caporaso10}
J~G Caporaso, J~Kuczynski, J~Stormbaugh, K~Bittinger, F~D Bushman, E~K
  Costello, and et~al.
\newblock Qiime allows analysis of high-throughput community sequencing data.
\newblock {\em Nat. Methods}, 7:335--336, 2010.

\bibitem{diag13}
{DIAG}.
\newblock Data intensive academic grid, June 2013.

\bibitem{edgar10}
R~C Edgar.
\newblock Search and clustering orders of magnitude faster than blast.
\newblock {\em Bioinformatics}, 26(19):2460--2461, 2010.

\bibitem{galvez10}
S~G\'{a}lvez, D~D\'{i}az, P~Hern\'{a}ndez, F~J Esteban, J~A Caballero, and
  G~Dorado.
\newblock Next-generation bioinformatics: using many-core processor
  architecture to develop a web service for sequence alignment.
\newblock {\em Bioinformatics}, 26(5):683--686, 2010.

\bibitem{gilles11}
A~Gilles, E~Megl\'{e}cz, N~Pech, S~Ferreira, T~Malausa, and J~F Martin.
\newblock Accuracy and quality assessment of 454 gs-flx titanium
  pyrosequencing.
\newblock {\em BMC Genomics}, 12:245--255, 2011.

\bibitem{goecks10}
J~Goecks, A~Nekrutenko, J~Taylor, and {The Galaxy Team}.
\newblock Galaxy: a comprehensive approach for supporting accessible,
  reproducible, and transparent computational research in the life sciences.
\newblock {\em Genome Biol.}, 11(8):R86, 2010.

\bibitem{haas11}
B~J Haas, D~Gevers, A~M Earl, M~Feldgarden, D~V Ward, and G~Giannoukos.
\newblock Chimeric 16s rrna sequence formation and detection in sanger and
  454-pyrosequenced pcr amplicons.
\newblock {\em Genome Res.}, 21:494--504, 2011.

\bibitem{heidrich12}
E~S Heidrich.
\newblock {\em Evaluation of microbial electrolysis cells in the treatment of
  domestic wastewater}.
\newblock PhD thesis, Newcastle University, May 2012.

\bibitem{heracle12}
{Heracle BioSoft S.R.L.}
\newblock Sff workbench, March 2012.

\bibitem{hull06}
D~Hull, K~Wolstencroft, R~Stevens, C~Goble, M~Pocock, P~Li, and T~Oinn.
\newblock Taverna: a tool for building and running workflows of services.
\newblock {\em Nucleic Acids Research}, 34:729--732, 2006.
\newblock Web Server Issue.

\bibitem{krebs89}
CJ~Krebs.
\newblock {\em Ecological Methodology}.
\newblock Harper \& Row, New York, 1989.

\bibitem{kuczynski11}
J~Kuczynski, J~Stombaugh, WA~Walters, A~Gonz{\'a}lez, JG~Caporaso, and
  R~Knight.
\newblock Using qiime to analyze 16s rrna gene sequences from microbial
  communities.
\newblock {\em Current protocols in bioinformatics / editoral board, Andreas D.
  Baxevanis ... {$[$}et al.{$]$}}, Chapter 10:Unit 10.7., 2011.

\bibitem{lin07}
Feng Lin, Heiko Schr\"oder, and Bertil Schmidt.
\newblock Solving the bottleneck problem in bioinformatics computing: An
  architectural perspective.
\newblock {\em The Journal of VLSI Signal Processing Systems for Signal, Image,
  and Video Technology}, 48(3):185--188, 2007.

\bibitem{liu12}
Y~Liu, B~Schmidt, and D~Maskell.
\newblock Cushaw: a cuda compatible short read aligner to large genomes based
  on the burrows-wheeler transform.
\newblock {\em Bioinformatics}, 2012.

\bibitem{lozupone05}
C~Lozupone and R~Knight.
\newblock Unifrac: a new phylogenetic method for comparing microbial
  communities.
\newblock {\em Appl. Environ. Mircobiol.}, 71(12):8228--8235, 2005.

\bibitem{matlab13}
MATLAB.
\newblock {\em version 8.1.0.604 (R2013a)}.
\newblock The MathWorks Inc., Natick, Massachusetts, 2013.

\bibitem{mcdonald12}
D~McDonald, M~N Price, J~Goodrich, E~P Nawrocki, T~Z DeSantis, A~Probst, G~L
  Andersen, R~Knight, and P~Hugenholtz.
\newblock An improved greengenes taxonomy with explicit ranks for ecological
  and evolutionary analyses of bacteria and archaea.
\newblock {\em ISME J.}, 6(3):610--618, 2012.

\bibitem{price10}
M~N Price, P~S Dehal, and A~P Arking.
\newblock Fasttree 2-approximately maximum-likelihood trees for large
  alignments.
\newblock {\em Plos One}, 5(3):e9490, 2010.

\bibitem{qiime13}
QIIME.
\newblock Denoising of 454 data sets.
\newblock http://qiime.org/tutorials/denoising\_454\_data.html.
\newblock Accessed: 2013-07-17.

\bibitem{anoise13}
Chrisopher Quince.
\newblock Software for pyrosequencing noise removal.
\newblock
  http://userweb.eng.gla.ac.uk/christopher.quince/Software/PyroNoise.html.
\newblock Accessed: 2013-07-17.

\bibitem{quince08}
Christopher Quince, Thomas~P Curtis, and William~T Sloan.
\newblock The rational exploration of microbial diversity.
\newblock {\em The ISME Journal}, 2:997--1006, 2008.

\bibitem{quince09}
Christopher Quince, Andrew Lanzen, Thomas~P Curtis, Russell~J Davenport, Neil
  Hall, Ian~M Head, Fiona~L Read, and William~T Sloan.
\newblock Accurate determination of microbial diversity from 454 pyrosequencing
  data.
\newblock {\em Nat. Methods}, 6:639--U27, 2009.

\bibitem{quince11}
Christopher Quince, Andrew Lanzen, Russell~J Davenport, and Peter~J Turbaugh.
\newblock Removing noise from pyrosequenced amplicons.
\newblock {\em BMC Bioinformatics}, 12:38, 2011.

\bibitem{ragan-kelley13}
Benjamin Ragan-Kelley, William~A Walters, Daniel McDonald, Justin Riley,
  Brian~B Granger, Antonio Gonzalez, Rob Knight, Fernando Perez, and J~Gregory
  Caporaso.
\newblock Collaborative cloud-enabled tools allow rapid, reproducible
  biological insights.
\newblock {\em The ISME Journal}, 7:461--464, 2013.

\bibitem{reeder10}
J~Reeder and R~Knight.
\newblock Rapidly denoising pyrosequencing amplicon reads by exploiting
  rank-abundance distributions.
\newblock {\em Nat. Methods}, 7:668--669, 2010.

\bibitem{schatz10}
Michael~C Schatz, Ben Langmead, and Steven~L Salzberg.
\newblock Cloud computing and the dna data race.
\newblock {\em Nat. Biotechnol.}, 28(7):691--693, 2010.

\bibitem{scholz12}
Matthew~B. Scholz, Chien-Chi Lo, and Patrick~SG Chain.
\newblock Next generation sequencing and bioinformatic bottlenecks: the current
  state of metagenomic data analysis.
\newblock {\em Current Opinion in Biotechnology}, 23(1):9--15, 2012.

\bibitem{tan06}
G~Tan, L~Xu, S~Feng, and N~Sun.
\newblock An experimental study of optimizing bioinformatics applications.
\newblock {\em Proceedings of IEEE International Parallel \& Distributed
  Processing Symposium (HiCOMB)}, pages 25--29, 2006.

\bibitem{wang07}
Q~Wang, G~M Garrity, J~M Tiedje, and J~R Cole.
\newblock Naive bayesian classifier for rapid assignment of rrna sequences into
  the new bacterial taxonomy.
\newblock {\em Appl. Environ. Microb.}, 73(16):5261--5267, 2007.

\bibitem{white11}
James~Robert White, Malcolm Matalka, W~Florian Fricke, and Samuel~V Angiuoli.
\newblock Cunningham: a blast runtime estimator.
\newblock {\em Nature Precedings}, January 2011.

\bibitem{wine13}
Wine.
\newblock Wine [version 1.5.25].
\newblock http://www.winehq.org, March 2013.

\end{thebibliography}
\bibliographystyle{plain}

\onecolumn
\begin{appendices}
\setcounter{table}{0}
\setcounter{figure}{0}
  \renewcommand\thetable{\thesection\arabic{table}}
  \renewcommand\thefigure{\thesection\arabic{figure}}
\section{Supplemental Material} \label{suppmat}
\begin{figure}[h]
\centering
\includegraphics[clip=true,width=4.5in]{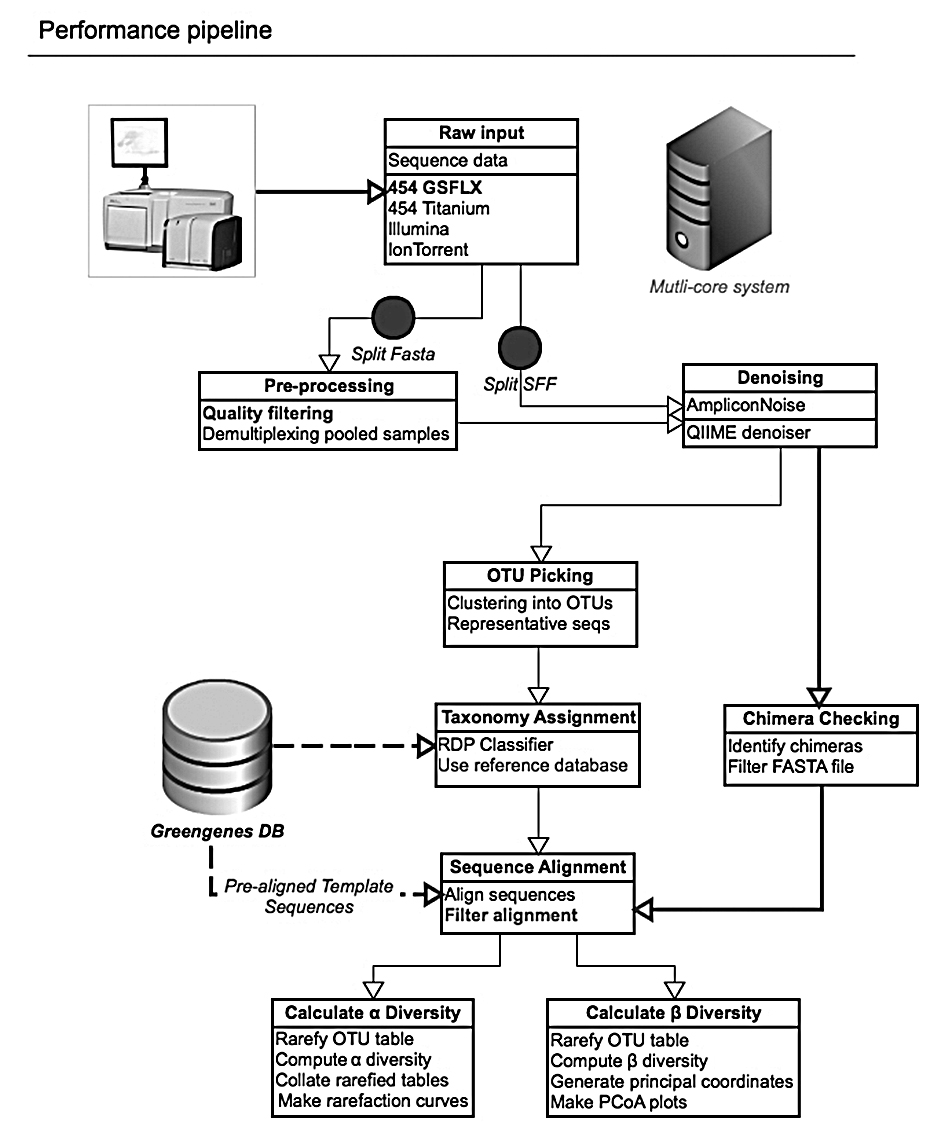}
\caption{The general analysis pipeline utilised within QIIME for modelling performance}\label{fig:PP}
\end{figure}

\begin{table}[!h]
\caption{Pipeline Parameters for Relevant Analysis Steps}
\label{tab:pipeparam}
\centering
\begin{tabular}{|c|c|c|c|}
\hline
Script                 & Parameters                                                                                     & Flag                                                      & Value                                                              \\ \hline \hline
Split libraries        & \begin{tabular}[c]{@{}c@{}}min seq length\\ max seq length\end{tabular}                        & \begin{tabular}[c]{@{}c@{}}-l\\ -L\end{tabular}           & \begin{tabular}[c]{@{}c@{}}150\\ 1000\end{tabular}                 \\ \hline
AmpliconNoise          & seqnoise resol.                                                                            & -s                                                        & \begin{tabular}[c]{@{}c@{}}25  (Titanium)\\ 30 (FLX)\end{tabular}  \\ \hline
Pick OTUs              & \begin{tabular}[c]{@{}c@{}}method\\ similarity\end{tabular}                                    & \begin{tabular}[c]{@{}c@{}}-m\\ -s\end{tabular}           & \begin{tabular}[c]{@{}c@{}}uclust\\ 0.97\end{tabular}              \\ \hline
\begin{tabular}[c]{@{}c@{}}Representative\\ set\end{tabular}      & \begin{tabular}[c]{@{}c@{}}method\\ sorting\end{tabular}                                       & \begin{tabular}[c]{@{}c@{}}-m\\ -s\end{tabular}           & \begin{tabular}[c]{@{}c@{}}first (clust. seed)\\ otu\end{tabular} \\ \hline
\begin{tabular}[c]{@{}c@{}}Taxonomy\\ assignment\end{tabular}     & \begin{tabular}[c]{@{}c@{}}method\\ confidence\end{tabular}                                    & \begin{tabular}[c]{@{}c@{}}-m\\ -c\end{tabular}           & \begin{tabular}[c]{@{}c@{}}rdp\\ 0.8\end{tabular}                  \\ \hline
Alignment              & \begin{tabular}[c]{@{}c@{}}method\\ pairwise method\\ min percent ID\\ min length\end{tabular} & \begin{tabular}[c]{@{}c@{}}-m\\ -a\\ -p\\ -e\end{tabular} & \begin{tabular}[c]{@{}c@{}}pynast\\ uclust\\ 75\\ 150\end{tabular} \\ \hline
Identify chimeras      & \begin{tabular}[c]{@{}c@{}}method\\ fragments\\ taxonomy depth\end{tabular}                    & \begin{tabular}[c]{@{}c@{}}-m\\ -n\\ -d\end{tabular}      & \begin{tabular}[c]{@{}c@{}}chimera slayer\\ 3\\ 4\end{tabular}     \\ \hline
Filter alignment       & \begin{tabular}[c]{@{}c@{}}allowed gap frac.\\ threshold\end{tabular}                       & \begin{tabular}[c]{@{}c@{}}-g\\ -t\end{tabular}           & \begin{tabular}[c]{@{}c@{}}0.999999\\ 3\end{tabular}               \\ \hline
\begin{tabular}[c]{@{}c@{}}Make phylo.\\ tree\end{tabular}  & method                                                                                         & -t                                                        & fasttree                                                           \\ \hline
\end{tabular}
\end{table}

\begin{table}[!h]
\caption{Pipeline Script Inputs According to Type. Representative Sequences are Equivalent to OTUs After Clustering}
\label{tab:seqtype}
\centering
      \begin{tabular}{|c|c|}
       \hline 
         Script & Input type\\ \hline\hline
        Quality filtering & Raw sequences\\ \hline
        Denoising & Trimmed sequences\\ \hline
        OTU picking & Denoised sequences\\ \hline
        Taxonomy assignment & Representative sequences\\ \hline
        Alignment & Representative sequences\\ \hline
        Chimera removal & Representative sequences\\ \hline
        Phylogeny & Chimera free OTUs\\ \hline
        OTU table & OTUs \& taxonomy assignments\\ \hline
        Diversity & OTU table \& phylogenetic tree\\ \hline
      \end{tabular}    
      \end{table}

\begin{figure}[!h]
\centering
\includegraphics[trim=0cm 0cm 0cm 0cm,clip=true,width=4.5in]{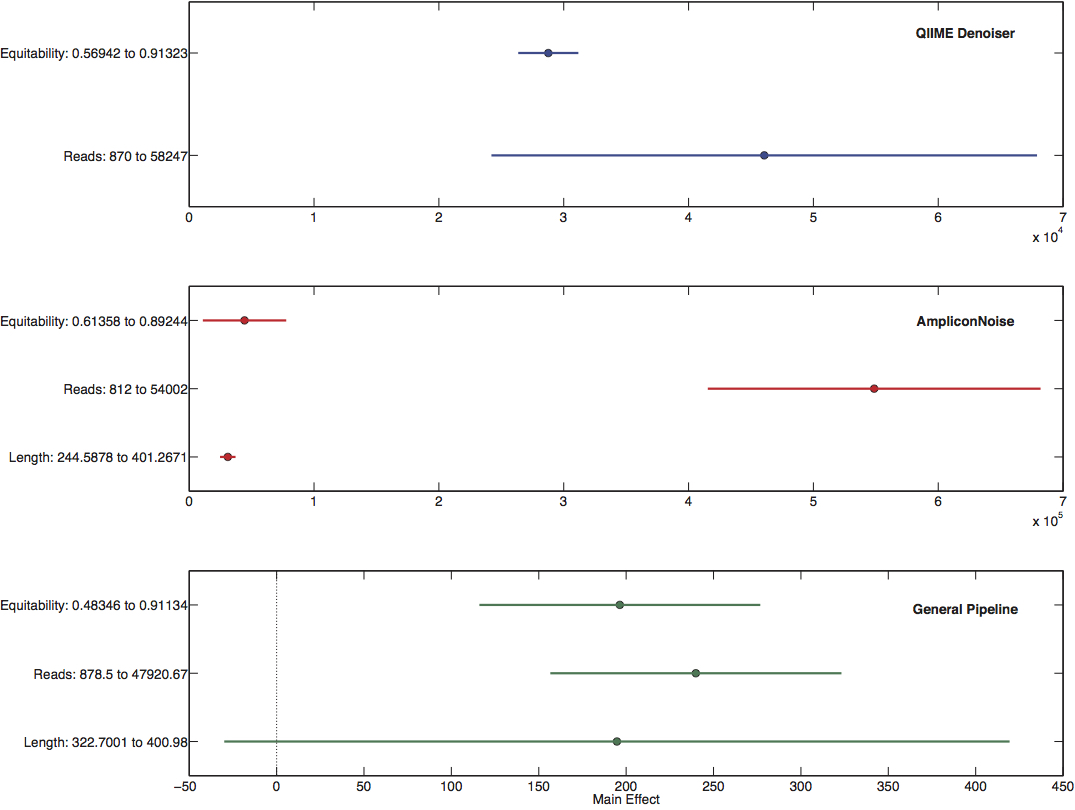}
\caption{Main effects plots for the three models}
\label{fig:maineff}
\end{figure}
\end{appendices}

\end{document}